\documentclass[amsmath,amssymb,twocolumn]{revtex4}
\usepackage[english]{babel}
\usepackage{braket}
\usepackage{gensymb}
\usepackage{graphicx}
\usepackage{courier}
\usepackage{amsmath}
\usepackage{textcomp}
\usepackage{xspace}
\usepackage{multirow}
\usepackage{bm}
\usepackage{soul}
\usepackage{amsfonts}
\usepackage{amssymb}
\usepackage{pifont}
\usepackage{bm}
\usepackage{xcolor}
\usepackage{hyperref}
\usepackage{mciteplus}

\newcommand\expect[3]{\langle #1 \vert #2 \vert #3 \rangle}
\newcommand\Expect[2]{\langle #1 \vert #2 \rangle}
\newcommand\av[1]{\left\langle #1 \right\rangle}

\renewcommand\b[1]{{\textbf{#1}}}
\newcommand\B[1]{{\bm #1}}

\newcommand\deriv[2]{\frac{\partial #1}{\partial #2}}

\begin{document}

\title{Time-dependent linear-response variational Monte Carlo}

\author{Bastien Mussard$^{1,2}$}\email{bastien.mussard@upmc.fr}
\author{Emanuele Coccia$^{1,3}$}\email{emanuele.coccia@nano.cnr.it}
\author{Roland Assaraf$^1$}
\author{Matt Otten$^4$}
\author{C. J. Umrigar$^4$}\email{cyrusumrigar@cornell.edu}
\author{Julien Toulouse$^1$}\email{julien.toulouse@upmc.fr}
\affiliation{
$^1$ Laboratoire de Chimie Th\'eorique, Universit\'e Pierre et Marie Curie, Sorbonne Universit\'es, CNRS, Paris, France\\
$^2$ Institut des Sciences du Calcul et des Donn\'ees, Universit\'e Pierre et Marie Curie, Sorbonne Universit\'es, Paris, France\\
$^3$ Dipartimento di Scienze Fisiche e Chimiche, Universit\'a degli Studi dell'Aquila, L'Aquila, Italy\\
$^4$ Laboratory of Atomic and Solid-State Physics, Cornell University, Ithaca, NY, USA}

\date{May 8, 2017}
\begin{abstract}
We present the extension of variational Monte Carlo (VMC) to the calculation of electronic excitation energies and oscillator strengths using
time-dependent linear-response theory. By exploiting the analogy existing between the linear method for wave-function optimisation and the generalised eigenvalue equation of linear-response theory, we formulate the equations of linear-response VMC (LR-VMC). This LR-VMC approach involves the first- and second-order derivatives of the wave function with respect to the parameters. We perform first tests of the LR-VMC method within the Tamm-Dancoff approximation using single-determinant Jastrow-Slater wave functions with different Slater basis sets on some singlet and triplet excitations of the beryllium atom. Comparison with reference experimental data and with configuration-interaction-singles (CIS) results shows that LR-VMC generally outperforms CIS for excitation energies and is thus a promising approach for calculating electronic excited-state properties of atoms and molecules.\\[0.1cm]

\end{abstract}

\maketitle
\section{Introduction}\label{intro}

Quantum Monte Carlo (QMC) methods \cite{bk:hammond,fou+01rmp,tou16} are a powerful and reliable alternative to
wave-function methods  and density-functional theory (DFT) for
quantum chemistry calculations, thanks to their favorable scaling with system size and to their suitability for high-performance computing infrastructures, such as petascale architectures.
Variational Monte Carlo (VMC) \cite{bres98} combines
Monte Carlo integration for computing the expectation value of the electronic
Hamiltonian $\hat{H}$ and the variational principle for the ground state.
VMC scales as $N^{3-4}$ (where $N$ is the number of electrons), similar to DFT scaling. The main
drawback of any QMC approach is the very large prefactor in the scaling, preventing the
systematic use of QMC in quantum chemistry calculations of medium- and large-size systems.
This drawback is alleviated by performing massive parallel calculations on supercomputers \cite{Coccia:2012kz,Coccia14}. 

A fundamental role is played by the trial wave function, often written as a product
of a determinantal part and a bosonic Jastrow factor \cite{dru+04prb} which depends on
interparticle distances (with electron-nucleus, electron-electron, higher many-body terms,\dots).
For example, one can use for the determinantal part a linear combination of
configuration state functions (CSF, i.e. spatial- and spin-symmetry adapted linear combinations of Slater determinants
of one-electron molecular orbitals) \cite{pet12}, or the antisymmetrised geminal power (AGP) ansatz (a single determinant
of geminal pairing functions \cite{cas+03jcp,cas+04jcp,zen14}). 
Furthermore, the optimisation of the wave function is crucial for an accurate description of both static and dynamic
electron correlation.
The linear method \cite{tou07,umr+07prl,tou08} allows one to
efficiently perform such an optimisation for all the parameters of the wave function, using only the first-order derivatives of the wave function with respect to the
parameters. 

The calculation of excited-state properties of
molecules (from prototypical models to complex organic dyes and biochromophores) still represents an
open challenge for theoreticians.
The two commonly used approaches are time-dependent density-functional theory, which
is not computationally demanding but sometimes lacks accuracy, 
and wave-function methods, which are more
accurate but very computationally demanding.
QMC methods were originally formulated for ground state problems and their extension to excited states is not straightforward.
Relatively few applications of QMC for electronic excitations are present in literature, see e.g. the singlet and triplet energies for the benchmark CH$_{2}$ diradical \cite{zim09}, the low-lying singlet excited states of biochromophores \cite{fili10,filippi2011bathochromic,val12,val13}, the $n \rightarrow \pi^*$ transition in acrolein \cite{tou12,flo14}, and the recent extension of the AGP ansatz for calculating excited-state energies \cite{zen+15jctc,dup15,zha16a,zha16b,neu16}.  

The basic idea of the present work stems from the formal
analogy existing between the linear method for wave function optimisation and time-dependent linear-response theory
\cite{bk:mc}. Indeed, the generalised eigenvalue equations of linear-response theory in the Tamm-Dancoff
approximation (TDA) and of the linear method at the ground-state minimum coincide.
Starting from this observation, we derive and implement the
linear-response equations in VMC (LR-VMC). This represents an extension of the well-established
time-dependent linear-response Hartree-Fock or multiconfiguration self-consistent
field methods, taking into account both static and dynamic electron correlations. 

The paper is organised as follows. In Section \ref{theory}, VMC and linear-response theory are briefly reviewed, and the LR-VMC method is presented and discussed in detail.
Results of LR-VMC calculations in the TDA of some singlet and triplet excitations of the beryllium atom are reported and discussed in Section \ref{res}.
Conclusions and perspectives for future work are given in Section \ref{con}.

\section{Theory}
\label{theory}

We first briefly review the form of the wave function that we use and the linear optimisation method. We then derive the time-dependent linear-response equations and show how to implement them in VMC.

\subsection{Wave-function parametrisation}

We consider Jastrow-Slater-type wave functions parametrised as \cite{tou07,tou08}
\begin{equation}
  \ket{\Psi(\b{p})} = \hat{J}(\bm{\alpha})  e^{\hat{\kappa}(\bm{\kappa})} \sum_{I=1}^{N_\text{CSF}}  c_I  \ket{C_I},
\end{equation}
where $\hat{J}(\B{\alpha})$ is a Jastrow factor operator depending on a set of parameters $\bm{\alpha}$, $e^{\hat{\kappa}(\bm{\kappa})}$ is the orbital rotation operator depending on a set of orbital rotation parameters $\bm{\kappa}$, and $\ket{C_I}$ are CSFs with associated coefficients $\b{c}=\{c_I\}$.
The CSFs are linear combinations of Slater determinants of orbitals $\ket{\phi_i}$, which are expanded in a basis of Slater functions $\{\ket{\chi_{\mu}}\}$
\begin{align}
& \ket{\phi_i} = \sum_{\mu=1}^{N_\text{basis}} \lambda_{i \mu} \ket{\chi_{\mu}}.
\end{align}
The Slater functions are centered on the nuclei and their spatial representation is
\begin{align}
& \langle \b{r} \ket{\chi_{\mu}}= N_n(\zeta)\;\; r^{n-1}\text{e}^{-\zeta r} \;\; Y_{\ell,m}(\theta,\phi),
\end{align}
each function being characterized by a set of quantum numbers $n,\ell,m$ and an exponent $\zeta$, $Y_{\ell,m}(\theta,\phi)$ are real spherical harmonics, and $N(\zeta)$ a normalisation factor. The full set of parameters to consider is thus $\b{p} = \left\{ \bm{\alpha}, \b{c}, \bm{\kappa}, \bm{\zeta} \right\}$ where $\bm{\zeta}$ stands for the set of exponents.

\subsection{Linear optimisation method}

The linear optimisation method\cite{tou07,umr+07prl,tou08} allows one to find the optimal parameters $\b{p}$ using an iterative procedure. At each iteration, we consider the intermediate-normalised wave function
\begin{equation}
\ket{\overline{\Psi}(\b{p})}
=\frac{\ket{\Psi(\b{p})}}{\Expect{\Psi_0}{\Psi(\b{p})}}
\label{eq:norm}
\end{equation}
where $\ket{\Psi_0} = \ket{\Psi(\b{p}^0)}$ is the wave function for the parameters $\b{p}^0$ at the current iteration (taken as normalised to unity, i.e. $\Expect{\Psi_0}{\Psi_0}=1$), and we expand it to linear order in the parameter variations $\bm{\Delta}\b{p} = \b{p} - \b{p}^0$,
\begin{equation}
  \ket{\overline{\Psi}_{\text{lin}}(\b{p})} = \ket{{\Psi}_{0}} + \sum_{i}  \Delta p_i \ket{\overline{\Psi}_i},
\label{}
\end{equation}
where $\ket{\overline{\Psi}_i}$ are the first-order derivatives of the wave function $\ket{\overline{\Psi}(\b{p})}$
\begin{equation}
\ket{\overline{\Psi}_i} = \left( \frac{\partial \ket{\overline{\Psi}(\b{p})}}{\partial p_i} \right)_{\b{p}=\b{p}^0} = \ket{\Psi_i} - \Expect{\Psi_0}{\Psi_i} \ket{\Psi_0},
\label{Psiib}
\end{equation}
where $\ket{{\Psi}_i} = \left( \partial \ket{\Psi(\b{p})}/\partial p_i \right)_{\b{p}=\b{p}^0}$ are the first-order derivatives of the original wave function $\ket{\Psi(\b{p})}$. Using the intermediate-normalised wave function has the advantage that the derivatives in Eq.~(\ref{Psiib}) are orthogonal to $\ket{\Psi_0}$, i.e. $\Expect{\Psi_0}{\overline{\Psi_i}}=0$. We then determine the parameter variations $\bm{\Delta}\b{p}$ by minimising the corresponding energy
\begin{equation}
E_\text{lin} = \min_{\b{p}}  \frac{\bra{\overline{\Psi}_{\text{lin}}(\b{p})} \hat{H} \ket{\overline{\Psi}_{\text{lin}}(\b{p})}}{\Expect{\overline{\Psi}_{\text{lin}}(\b{p})}{\overline{\Psi}_{\text{lin}}(\b{p})}},
\label{Elinmin}
\end{equation}
we update the parameters as $\b{p}^0\to \b{p}^0 + \bm{\Delta} \b{p}$, and iterate until convergence.

The minimisation in Eq.~(\ref{Elinmin}) leads to the following generalized eigenvalue equation to be solved at each iteration
\begin{align}
\begin{pmatrix}
  E_0 & \b{g}_\text{R}^\text{T}/2 \\
  \b{g}_\text{L}/2 & \b{H}
\end{pmatrix}
\begin{pmatrix}
  1 \\ \bm{\Delta} \b{p}
\end{pmatrix}
= E_\text{lin}
\begin{pmatrix}
  1 & \b{0}^\text{T} \\
  \b{0} & \b{S}
\end{pmatrix}
\begin{pmatrix}
  1 \\ \bm{\Delta} \b{p}
\end{pmatrix},
\label{eq:linmethod}
\end{align}
where $E_0 = \bra{\Psi_0} \hat{H} \ket{\Psi_0}$ is the current energy, $g_{\text{L},i}=2 \bra{\overline{\Psi}_i} \hat{H} \ket{\Psi_0}$ and $g_{\text{R},j}=2 \bra{{\Psi}_0} \hat{H} \ket{\overline{\Psi}_j}$ are the left and right energy gradients (identical except on a finite Monte Carlo sample), and $H_{ij} = \bra{\overline{\Psi}_i} \hat{H} \ket{\overline{\Psi}_j}$ is the Hamiltonian matrix in the basis of the first-order wave function derivatives, and $S_{ij} = \Expect{\overline{\Psi}_i}{\overline{\Psi}_j}$ is the overlap matrix in this basis. Note that in Eq.~(\ref{eq:linmethod}), $\b{0}$ and $\b{0}^\text{T}$ stand for the zero column vector and the zero row vector, respectively.

\subsection{Linear-response theory}
\label{linearresponse}

Starting from the previously optimised wave function, we introduce now a time-dependent perturbation (e.g, interaction with an electric field) in the Hamiltonian
\begin{equation}
\hat{H}(t) = \hat{H} + \gamma \hat{V}(t),
\end{equation}
where $\gamma$ is a coupling constant. The approximate ground-state wave function $\ket{\Psi(\b{p}(t))}$ evolves in time through its parameters $\b{p}(t)$, which become now generally complex. As before, it is convenient to introduce the intermediate-normalised wave function
\begin{equation}
\ket{\overline{\Psi}(\b{p}(t))}
=\frac{\ket{\Psi(\b{p}(t))}}{\Expect{\Psi_0}{\Psi(\b{p}(t))}},
\end{equation}
where $\ket{\Psi_0} = \ket{\Psi(\b{p}^0)}$ is the wave function for the initial parameters $\b{p}^0$, again taken as normalised to unity (i.e., $\Expect{\Psi_0}{\Psi_0}=1$). At each time, the time-dependent parameters $\b{p}(t)$ can be found from the Dirac-Frenkel variational principle (see, e.g., Ref.\citenum{bk:mc})
\begin{equation}
\deriv{}{p_i^*}
\frac{\expect{\overline{\Psi}(\b{p}(t)) } { \hat{H}(t) - i \frac{\partial} {\partial t } } { \overline{\Psi}(\b{p}(t)) }}
     {\Expect{\overline{\Psi}(\b{p}(t)) } { \overline{\Psi}(\b{p}(t))  }}
     = 0.
\label{eq:df}
\end{equation}
To apply Eq.~(\ref{eq:df}) in linear order in $\gamma$, we start by expanding the wave function $\ket{\overline{\Psi}(\b{p}(t))}$ around $\b{p}^0$ to second order in the parameter variations $\Delta  \b{p}(t) = \b{p}(t) - \b{p}^{0}$
\begin{eqnarray}
  \ket{\overline{\Psi}(\b{p}(t))}
&=&
\ket{\Psi_0} + \sum_{i} \Delta p_{i}(t) \ket{\overline{\Psi}_i}
\nonumber\\
&&+\frac {1} {2} \sum_{i,j} \Delta p_{i}(t) \Delta p_{j}(t)  \ket{\overline{\Psi}_{ij}} + \cdots,
\label{PsiExpand}
\end{eqnarray}
where $\ket{\overline{\Psi}_i}$ are the first-order derivatives of $\ket{\overline{\Psi}(\b{p})}$ already introduced in Eq.~(\ref{Psiib}), and $\ket{\overline{\Psi}_{ij}}$ are the second-order derivatives of the wave function $\ket{\overline{\Psi}(\b{p})}$
\begin{eqnarray}
\ket{\overline{\Psi}_{ij}} &=& \left( \frac{\partial^2 \ket{\overline{\Psi}(\b{p})}}{\partial p_i\partial p_j} \right)_{\b{p}=\b{p}^0}
\nonumber \\
&=&\ket{\Psi_{ij}}
-\Expect{\Psi_0}{\Psi_j}
 \ket{\Psi_i}
-\Expect{\Psi_0}{\Psi_i}
 \ket{\Psi_j}
\nonumber \\
&&+\left(
2\Expect{\Psi_0}{\Psi_i}
 \Expect{\Psi_0}{\Psi_j}
-\Expect{\Psi_0}{\Psi_{ij}}
 \right)\ket{\Psi_0},
\label{Psibij2}
\end{eqnarray}
where $\ket{{\Psi}_{ij}} = \left( \partial^2 \ket{\Psi(\b{p})}/\partial p_i\partial p_j \right)_{\b{p}=\b{p}^0}$ are the second-order derivatives of the original wave function $\ket{\Psi(\b{p})}$. Again, the advantage of using the intermediate-normalised wave function is that the second-order derivatives are orthogonal to $\ket{\Psi_0}$, i.e. $\Expect{\Psi_0}{\overline{\Psi}_{ij}}=0$. Plugging Eq.~(\ref{PsiExpand}) into Eq.~(\ref{eq:df}) and keeping only first-order terms in $\bm{\Delta} \b{p}(t)$, in the limit of a vanishing perturbation ($\gamma \rightarrow 0$), we find
\begin{equation}
\b{A} \; \bm{\Delta} \b{p} (t)  + \b{B} \; \bm{\Delta} \b{p} (t)^* = i \b{S} \; \frac{\partial \bm{\Delta} \b{p} (t)}{\partial t},
\end{equation}
with the matrices $A_{ij}= \bra{\overline{\Psi}_i} \hat{H} - E_0 \ket{\overline{\Psi}_j}= H_{ij} - E_0 S_{ij}$ where $E_0$ is the ground-state energy, $B_{ij}= \bra{\overline{\Psi}_{ij}} \hat{H}\ket{\Psi_0}$, and $S_{ij} = \Expect{\overline{\Psi}_i}{\overline{\Psi}_j}$. If we look for free-oscillation solutions of the form
 \begin{equation}
 \Delta  \b{p}(t) =  \b{X}  e^{-i\omega_n t}  + \b{Y}^{*}  e^{i\omega_n t},
 \end{equation}
where $\omega_n$ corresponds to an excitation (or de-excitation) energy, we arrive at the linear-response equation in the form of a non-Hermitian generalized eigenvalue equation\cite{bk:mc}
 \begin{equation}
   \begin{pmatrix}
     \b{A}     & \b{B}     \\
     \b{B}^{*} & \b{A}^{*} \\
   \end{pmatrix}
   \begin{pmatrix}
     \b{X}_n \\ \b{Y}_n
   \end{pmatrix}
   = \omega_{n}
   \begin{pmatrix}
     \b{S}   &  \b{0}       \\
     \b{0}   & -\b{S}^{*} \\
   \end{pmatrix}
   \begin{pmatrix}
     \b{X}_n \\ \b{Y}_n
   \end{pmatrix}.
\label{eq:main}
\end{equation}

The Tamm-Dancoff approximation (TDA) corresponds to neglecting the contributions from $\b{B}$, leading to
\begin{equation}
\b{A} \b{X}_n = \omega_n \b{S} \b{X}_n.
\label{eq:tda}
\end{equation}
At the ground-state minimum,  i.e. when the energy gradient is zero, the generalised eigenvalue equation of the linear method in Eq.~(\ref{eq:linmethod}) is equivalent to the TDA equation~(\ref{eq:tda}) which directly gives excitation energies $\omega_n = E_\text{lin} - E_0$.

Finally, the oscillator strength $f_{n}$ for the transition from the ground state to the excited state $n$ (with excitation energy $\omega_n$) can be easily extracted from the response vector $(\b{X}_n,\b{Y}_n)$
\begin{equation}
f_{n} = \frac{2}{3} \omega_{n} \sum_{\alpha=x,y,z} \left [ (\b{X}_{n} + \b{Y}_{n})^{\text{T}} \bm{\mu}^{\alpha} \right ]^{2},
\end{equation}
where $\bm{\mu}^{\alpha}$ is the vector containing the transition dipole moments for the component $\alpha$ ($x$, $y$, or $z$) between the ground-state wave function $\ket{\Psi_0}$ and the wave-function derivative $\ket{\overline{\Psi}_i}$
\begin{align}
\mu^{\alpha}_{i}
&=\expect{ \overline{\Psi}_{i} }{ \hat{\mu}^{\alpha} }{ \Psi_{0} },
\end{align}
and $\hat{\mu}^{\alpha}$ is the electronic dipole operator.

\subsection{Realisation in VMC}

We now give the expressions for performing linear-response calculations in VMC, referred to as LR-VMC, i.e. for calculating the expressions in Section~\ref{linearresponse} in a VMC run. For convenience, we also recall the expressions necessary for the linear optimisation method.

The current ground-state energy is calculated as
\begin{align}
E_0 = \av{E_\text{L}(\b{R})},
\end{align}
where $E_\text{L}(\b{R}) = [H\Psi_0(\b{R})]/\Psi_0(\b{R})$ is the local energy and $\av{...}$ stands for an average on a finite Monte Carlo sample of points $\b{R}_k$ distributed according to $\Psi_0(\b{R})^2$, with $\b{R}=(\b{r}_1,\b{r}_2,...,\b{r}_N)$ designating the electron coordinates. The left and right energy gradients are evaluated as
\begin{eqnarray}
g_{\text{L},i}&=&2 \av{\frac{\overline{\Psi}_i(\b{R})}{\Psi_0(\b{R})} \frac{H\Psi_0(\b{R})}{\Psi_0(\b{R})}}
\nonumber\\
&=& 2 \Biggl[ \av{\frac{{\Psi}_i(\b{R})}{\Psi_0(\b{R})} E_\text{L}(\b{R})} - \av{\frac{{\Psi}_i(\b{R})}{\Psi_0(\b{R})}} \av{E_\text{L}(\b{R})}\Biggl],
\nonumber\\
\label{gLi}
\end{eqnarray}
and
\begin{eqnarray}
g_{\text{R},j}&=&2 \av{\frac{{\Psi}_0(\b{R})}{\Psi_0(\b{R})} \frac{H\overline{\Psi}_j(\b{R})}{\Psi_0(\b{R})}}
\nonumber\\
&=& 2 \Biggl[ \av{\frac{{\Psi}_j(\b{R})}{\Psi_0(\b{R})} E_\text{L}(\b{R})} - \av{\frac{{\Psi}_j(\b{R})}{\Psi_0(\b{R})}} \av{E_\text{L}(\b{R})} 
\nonumber\\
&&+ \av{E_{\text{L},j}(\b{R})} \Biggl],
\label{gRj}
\end{eqnarray}
where $E_{\text{L},j}(\b{R})$ is the first-order derivative of the local energy
\begin{equation}
E_{\text{L},j}(\b{R})=\frac{H\Psi_j(\b{R})}{\Psi_0(\b{R})} - \frac{\Psi_j(\b{R})}{\Psi_0(\b{R})} E_\text{L}(\b{R}).
\end{equation}
Note that, in the limit of an infinite sample, $\av{E_{\text{L},j}(\b{R})}=0$ due to the hermiticity of the Hamiltonian, and therefore the left and right gradients become identical.

\begin{widetext}
The elements of the overlap matrix $\b{S}$ are calculated as
\begin{eqnarray}
  S_{ij}&=&
  \av{\frac{\overline{\Psi}_i(\b{R})}{\Psi_0(\b{R})}\frac{\overline{\Psi}_j(\b{R})}{\Psi_0(\b{R})}}
= \av{\frac{\Psi_i(\b{R})}{\Psi_0(\b{R})}\frac{\Psi_j(\b{R})}{\Psi_0(\b{R})}}
  -\av{\frac{\Psi_i(\b{R})}{\Psi_0(\b{R})}}\av{\frac{\Psi_j(\b{R})}{\Psi_0(\b{R})}},
\end{eqnarray}
and the elements of the matrix $\b{H}$ are evaluated as
\begin{eqnarray}
\label{Hij}
  H_{ij}
  &=&
  \av{\frac{\overline{\Psi}_i(\b{R})}{\Psi_0(\b{R})}   \frac{H\overline{\Psi}_j(\b{R})}{\Psi_0(\b{R})} }
\nonumber\\
&=&
   \av{\frac{\Psi_i(\b{R})}{\Psi_0(\b{R})}\frac{\Psi_j(\b{R})}{\Psi_0(\b{R})}E_\text{L}(\b{R})}
  -\av{\frac{\Psi_i(\b{R})}{\Psi_0(\b{R})}}\av{\frac{\Psi_j(\b{R})}{\Psi_0(\b{R})}E_\text{L}(\b{R})}
  -\av{\frac{\Psi_j(\b{R})}{\Psi_0(\b{R})}}\av{\frac{\Psi_i(\b{R})}{\Psi_0(\b{R})}E_\text{L}(\b{R})}
\nonumber\\
&&\quad  +\av{\frac{\Psi_i(\b{R})}{\Psi_0(\b{R})}E_{\text{L},j}(\b{R})}
  -\av{\frac{\Psi_i(\b{R})}{\Psi_0(\b{R})}}\av{E_{\text{L},j}(\b{R})}
  +\av{\frac{\Psi_i(\b{R})}{\Psi_0(\b{R})}}\av{\frac{\Psi_j(\b{R})}{\Psi_0(\b{R})}}\av{E_\text{L}(\b{R})}.
\end{eqnarray}

The elements of the matrix $\b{A}$ are then given by
\begin{eqnarray}
  A_{ij} &=& H_{ij} - E_0 S_{ij}
\nonumber\\
&=&
   \av{\frac{\Psi_i(\b{R})}{\Psi_0(\b{R})}\frac{\Psi_j(\b{R})}{\Psi_0(\b{R})}E_\text{L}(\b{R})}
  -\av{\frac{\Psi_i(\b{R})}{\Psi_0(\b{R})}}\av{\frac{\Psi_j(\b{R})}{\Psi_0(\b{R})}E_\text{L}(\b{R})}
  -\av{\frac{\Psi_j(\b{R})}{\Psi_0(\b{R})}}\av{\frac{\Psi_i(\b{R})}{\Psi_0(\b{R})}E_\text{L}(\b{R})}
\nonumber\\
&&\quad  +\av{\frac{\Psi_i(\b{R})}{\Psi_0(\b{R})}E_{\text{L},j}(\b{R})}
  -\av{\frac{\Psi_i(\b{R})}{\Psi_0(\b{R})}}\av{E_{\text{L},j}(\b{R})}
\nonumber\\
&&\quad
  -\av{\frac{\Psi_i(\b{R})}{\Psi_0(\b{R})}\frac{\Psi_j(\b{R})}{\Psi_0(\b{R})}}\av{E_\text{L}(\b{R})}
  +2\av{\frac{\Psi_i(\b{R})}{\Psi_0(\b{R})}}\av{\frac{\Psi_j(\b{R})}{\Psi_0(\b{R})}}\av{E_\text{L}(\b{R})},
\label{Aij}
\end{eqnarray}
and the elements of the matrix $\b{B}$ are
\begin{eqnarray}
  B_{ij}
  &=&
 \av{\frac{\overline{\Psi}_{ij}(\b{R})}{\Psi_0(\b{R})} \frac{H\Psi_0(\b{R})}{\Psi_0(\b{R})}}
\nonumber\\
&=&
 \av{\frac{\Psi_{ij}(\b{R})}{\Psi_0(\b{R})}E_\text{L}(\b{R})}
 - \av{\frac{\Psi_{ij}(\b{R})}{\Psi_0(\b{R})}} \av{E_\text{L}(\b{R})}
\nonumber\\
&&\quad
 -\av{\frac{\Psi_i(\b{R})}{\Psi_0(\b{R})}} \av{\frac{\Psi_{j}(\b{R})}{\Psi_0(\b{R})}E_\text{L}(\b{R})}
 -\av{\frac{\Psi_j(\b{R})}{\Psi_0(\b{R})}} \av{\frac{\Psi_{i}(\b{R})}{\Psi_0(\b{R})}E_\text{L}(\b{R})}
 +2\av{\frac{\Psi_i(\b{R})}{\Psi_0(\b{R})}}
         \av{\frac{\Psi_j(\b{R})}{\Psi_0(\b{R})}}
         \av{E_\text{L}(\b{R})}.
\end{eqnarray}
\end{widetext}

Finally, the expression of the transition dipole moment needed for calculating oscillator strengths is
\begin{eqnarray}
\mu^{\alpha}_{i} &=& \av{\frac{\overline{\Psi}_i(\b{R})}{\Psi_0(\b{R})} \mu^\alpha(\b{R})}
\nonumber\\
&=& \av{\frac{{\Psi}_i(\b{R})}{\Psi_0(\b{R})} \mu^\alpha(\b{R})} - \av{\frac{{\Psi}_i(\b{R})}{\Psi_0(\b{R})}} \av{\mu^\alpha(\b{R})},
\label{mu}
\end{eqnarray}
where $\mu^\alpha(\b{R}) = - \sum_{k=1}^N r_{k,\alpha}$ is the $\alpha$-component of electronic dipole moment.

In the linear optimisation method, using the non-symmetric estimator of the matrix $\b{H}$ in Eq.~(\ref{Hij}) instead of a symmetrised one has the advantage of leading to the strong zero-variance principle of Nightingale and Melik-Alaverdian \cite{zero}: in the limit where the current wave function $\ket{\Psi_0}$ and its first-order derivatives $\ket{\overline{\Psi}_i}$ form a complete basis for expanding an exact eigenstate of the Hamiltonian, the parameter variations $\bm{\Delta}\b{p}$ and the associated energy $E_\text{lin}$ are found from Eq.~(\ref{eq:linmethod}) with zero variance provided that the Monte Carlo sample size is larger than the number of parameters (see discussion in Ref.~\citenum{tou08}).
Unfortunately, this strong zero-variance principle does not apply when solving the linear-response equation~(\ref{eq:main}). However, in the limit where $\ket{\Psi_0}$ is an exact eigenstate of the Hamiltonian, the left energy gradient $g_{\text{L},i}$ in Eq.~(\ref{gLi}) vanishes with zero variance, and thus the TDA linear equation~(\ref{eq:tda}) becomes equivalent to Eq.~(\ref{eq:linmethod}) for calculating excited-state energies even on a finite Monte Carlo sample. Therefore, in this case, the strong zero-variance principle applies to the calculation of the response vectors $\b{X}_n$ and excitation energies $\omega_n$.

\subsection{Computational details}

The calculations shown here were performed using the QMC program CHAMP~\cite{CHAMP}, starting from Hartree-Fock calculations done with GAMESS~\cite{SchBalBoaElbGorJenKosMatNguSuWinDupMon-JCC-93}. Two Slater basis sets of different sizes were used, namely the VB1 and VB2 basis set from Ref. \citenum{ema03}. The VB1 basis set has five $s$ and one $p$ Slater functions ($[5s,1p]$), whereas the VB2 basis set has six $s$, two $p$, and one $d$ Slater functions ($[6s,2p,1d]$). We use a flexible Jastrow factor consisting of the exponential of the sum of electron-nucleus, electron-electron and electron-electron-nucleus terms, written as systematic polynomial and Pad\'e expansions~\cite{Umr-UNP-XX,FilUmr-JCP-96,GucSanUmrJai-PRB-05}, with 4 electron-nucleus parameters, 5 electron-electron parameters and 15 electron-electron-nucleus parameters. For each VMC calculation, 10$^{4}$ blocks were employed with 10$^{4}$ steps each. One block was used for equilibration of the VMC distribution. 

\section{Results}
\label{res}

\begin{figure*}
\begin{center}
\includegraphics[width=.49\linewidth]{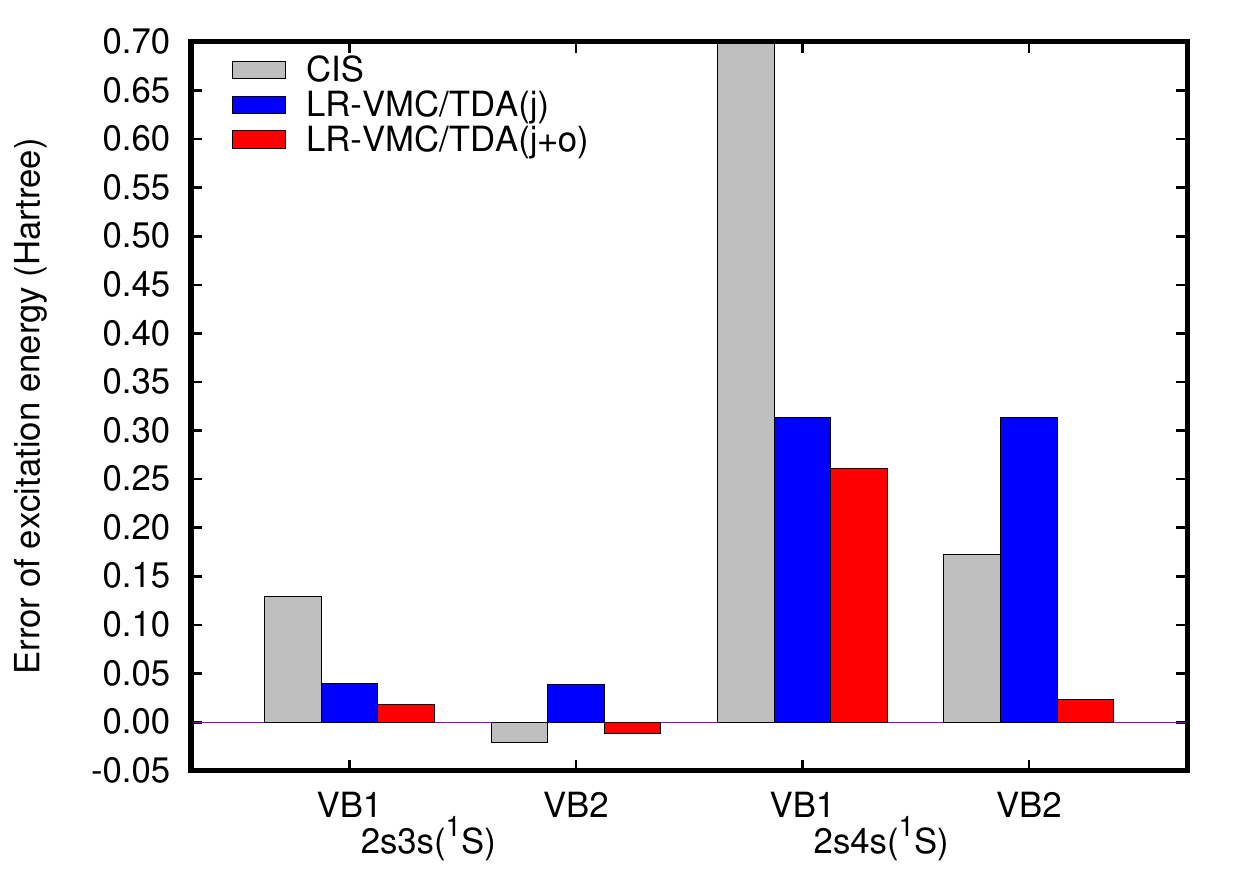}
\includegraphics[width=.49\linewidth]{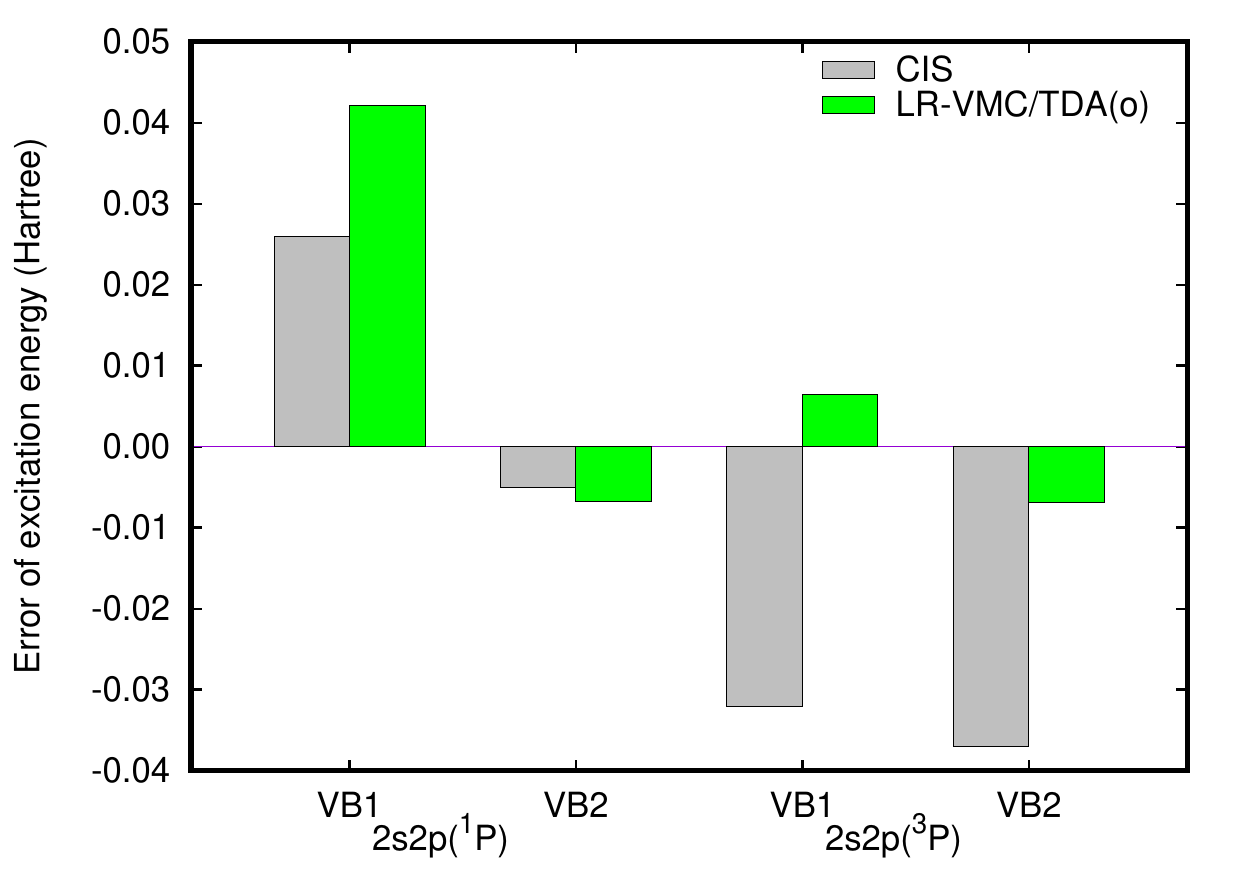}
\end{center}
\caption{Errors with respect to experimental values \cite{kra97} of CIS (grey), and LR-VMC/TDA excitation energies including the response of the Jastrow parameters (j) (blue) and the response of the Jastrow and orbitals parameters (j+o) (red) for $S$ excited states (left), and the response of the orbital parameters (o) (green) for $P$ excited states (right). These results are also detailed in the tables.}
\label{fig:all2}
\end{figure*}

The beryllium atom was used as a first test of the LR-VMC approach, since accurate experimental reference values for the excitation energies are available from Ref. \citenum{kra97}. An accurate description of the Be ground state requires a multiconfigurational wave function for accounting for the near-degeneracy between the 2$s$ and 2$p$ orbitals. However, for these preliminary tests, we present only results of calculations using a Jastrow-Slater single-determinant wave function for the ground state using TDA linear-response theory. This choice is motivated by the fact that a direct comparison between the LR-VMC/TDA method and configuration-interaction-singles (CIS) calculations represents a simple but essential first step for validating our approach.
We expect LR-VMC/TDA to outperform CIS because the Jastrow factor in LR-VMC should account for a substantial part of electronic correlation,
and we find this to be the case for most of the excitations studied.
The results are presented both as errors with respect to the experimental values in Figure \ref{fig:all2} and as detailed excitation energies in the subsequent tables.

In Table \ref{tab:2s3s} results for the singlet 2$s$3$s$ ($^{1}S$) state are reported. The effect of the Slater basis set adopted is dramatic at the CIS level, as a reasonable agreement with the reference experimental value of 0.249 Hartree is found only when the VB2 basis set is used. LR-VMC/TDA values are labelled as follow: (j) designates the response of the Jastrow parameters only, while (j+o) is the response of both the Jastrow and orbital parameters. The response of the Jastrow factor substantially improves upon the CIS VB1 estimate, going from 0.378 to 0.2888(1) Hartree.
The excitation energy improves further when the response of the orbital parameters are included in the LR-VMC/TDA calculation, yielding an error of around 0.02 Hartree with respect to the experimental value. 
Increasing the size of the Slater basis set, i.e. moving from VB1 to VB2, we obtain a fair agreement with the experimental data when both the Jastrow and the orbital parameters are included in the response (0.2378(2) Hartree).

\begin{table}
  \caption{\label{tab:2s3s} $2s3s$ ($^{1}S$) excitation energies (in Hartree) for the beryllium atom calculated using CIS and LR-VMC/TDA including the response of the Jastrow parameters (j) and of the Jastrow and orbital parameters (j+o). The experimental value is taken from Ref. \citenum{kra97}.}
   \begin{tabular}{c | c | c | c  || c  }
        & CIS & LR-VMC/TDA(j) & LR-VMC/TDA(j+o)  & Exp. \\
   \hline\hline
     VB1        &   0.378   & 0.2888(1) &   0.2672(1) &  0.249 \\
     VB2       &   0.228   & 0.2880(1) &   0.2378(2)  &  0.249 \\
  \end{tabular}
\end{table}

\begin{table}
 \caption{\label{tab:2s4s} $2s4s$ ($^{1}S$) excitation energies (in Hartree) for the beryllium atom calculated using CIS and LR-VMC/TDA including the response of the Jastrow parameters (j) and of the Jastrow and orbital parameters (j+o). The experimental value is taken from Ref. \citenum{kra97}.}
   \begin{tabular}{c | c | c | c  || c  }
       & CIS & LR-VMC/TDA(j) & LR-VMC/TDA(j+o) &  Exp. \\
   \hline\hline
    VB1        &   2.639   &  0.6106(1)  &    0.5578(2) &  0.297  \\
    VB2        &   0.470   &  0.6104(1)  &    0.321(3)   &  0.297  \\
  \end{tabular}
\end{table}

The singlet 2$s$4$s$ ($^{1}S$) excitation is higher in energy, and CIS fails to recover the experimental result of 0.297 Hartree, for both basis sets, as shown in Table \ref{tab:2s4s}.
As already mentioned for the 2$s$3$s$ excitation, the response of the Jastrow factor plays an important role for the VB1 basis set, reducing the error in the excitation energy by around 2 Hartree.
Including the orbital parameters in the response lowers the excitation energy further to 0.5578(2), but this is still a large overestimate of the experimental value.
With the VB2 basis, the LR-VMC/TDA(j+o) calculation outperforms CIS, but a substantial error ($>$0.02 Hartree) still remains for this high-lying excitation.
The failure of VB1 and, to a lesser extent, of VB2 is likely due to the poor description of the 4$s$ orbital.

The extension of our proposed approach to $P$ excitations is straightforward, with a relaxation of the spatial symmetry constraints in the orbital rotation parameters. Note that the Jastrow factor employed in this work only depends on interparticle distances, i.e. it has spherical symmetry, and therefore excited states with $P$ symmetry cannot be represented with the wave-function derivatives with respect to the Jastrow parameters. For this reason, only results concerning the response of the orbitals (o) are reported for the $P$ excitations. In Table \ref{tab:2s2p}, results for the singlet 2$s$2$p$ ($^{1}P$) state are given, which is the lowest energy excitation in the beryllium atom. The CIS calculations with the VB1 and VB2 basis sets show a fair agreement with the reference value of 0.194 Hartree, the CIS calculation using the VB2 basis set being only 5 mHartree below it. 
The LR-VMC/TDA(o) estimate is also close to the experimental reference when the VB2 basis set is employed (0.1873(2) Hartree), while for the VB1 basis set LR-VMC/TDA(o) greatly overestimates the excitation energy. 

\begin{table}
  \caption{\label{tab:2s2p} $2s2p$ ($^{1}P$) excitation energies (in Hartree) for the beryllium atom calculated using CIS and LR-VMC/TDA including the response of the orbital parameters (o). The experimental value is taken from Ref. \citenum{kra97}.}
   \begin{tabular}{c | c | c || c }
       & CIS &  LR-VMC/TDA(o)  & Exp.  \\
   \hline\hline
    VB1        &   0.220     &  0.2358(1) & 0.194 \\
    VB2        &   0.189     &  0.1873(2) & 0.194 \\
  \end{tabular}
\end{table}
\begin{table}
   \caption{\label{tab:2s2pt} $2s2p$ ($^{3}P$) excitation energies (in Hartree) for the beryllium atom calculated using CIS and LR-VMC/TDA including the response of the  orbital parameters (o). The experimental value is taken from Ref. \citenum{kra97}.}
   \begin{tabular}{c |  c | c || c }
       & CIS &  LR-VMC/TDA(o)  & Exp. \\
   \hline\hline
    VB1        &   0.068     &  0.1064(1) & 0.100 \\
    VB2        &   0.063     &  0.0929(2) & 0.100 \\
  \end{tabular}
\end{table}
\begin{table}
   \caption{\label{tab:fosc} Oscillator strength $f$ corresponding to the $2s2p$  ($^{1}P$) excitation of the beryllium atom computed using CIS and LR-VMC/TDA including the response of the orbital parameters (o). The experimental value is taken from Ref. \citenum{SchKoc-PRA-00}.}
   \begin{tabular}{c | c |  c || c }
        & CIS   & LR-VMC/TDA(o)     & Exp. \\
   \hline\hline
    VB1  & 0.648 & 0.435(1)& 1.34(3)\\
    VB2  & 0.669 & 0.57(2) & 1.34(3)\\
  \end{tabular}
\end{table}
Similarly, our implementation of linear response allows us to easily compute triplet excitations by considering triplet orbital rotation parameters. The CIS calculation underestimates the correct excitation energy by more than 30 mHartree, while the LR-VMC/TDA(o) excitation energies are very close to the reference values of 0.100 Hartree. The basis set effects are small is this case. \\
Finally, we computed the oscillator strength $f$ (Table \ref{tab:fosc}) corresponding to the singlet $2s2p$ ($^{1}P$) excitation, which is non zero according to selection rules. The LR-VMC(o) oscillator strengths seem more sensitive to the basis set compared to the CIS oscillator strengths. Moreover, the inclusion of the Jastrow factor does not improve the oscillator strength.

\section{Conclusions and Perspectives}
\label{con}
In this work we have presented a formulation of time-dependent linear-response theory in the VMC framework using a Jastrow-Slater wave function. Compared to state-specific or state-average excited-state QMC methods, the advantage of this LR-VMC approach is that, after optimizing only one ground-state wave function, one can easily calculate several excitation energies of different spatial or spin symmetry. Compared to similar linear-response quantum chemistry methods, the presence of the Jastrow factor in LR-VMC allows one to explicitly treat a part of dynamical correlation.
A disadvantage of the method is that the excitation energies are much more sensitive than the ground-state energy to the quality of the optimized ground-state wave function.
This is true in other linear-response quantum-chemistry methods as well, but is a bigger drawback in a method that employs stochastic optimization.

Using a Jastrow-Slater single-determinant wave function and the TDA, the LR-VMC method was shown to be more accurate that CIS for most of the excitation energies of the beryllium atom that were studied. The LR-VMC approach thus seems a promising method for calculating electronic excitation energies. In the near future, a systematic study on a set of molecules will be an essential step to further validate the proposed methodology, together with calculations using the full response equation beyond the TDA. Also, we will explore using multideterminant wave functions, larger basis sets, and including the wave-function derivatives with respect to the exponents of the Slater functions.

\section*{Acknowledgments}

EC thanks University of L'Aquila for financial support and the Laboratoire de Chimie Th\'eorique for computational resources.
MO and CJU were supported in part by NSF grant ACI-1534965.

\bibliographystyle{ieeetr}

\end{document}